# An analytic investigation for the edge effect on mechanical properties of graphene nanoribbons

Guang-Rong Han, Jia-Sheng Sun, and Jin-Wu Jiang

# An analytic investigation for the edge effect on mechanical properties of graphene nanoribbons

Guang-Rong Han, Jia-Sheng Sun, and Jin-Wu Jiang[a)]
*Shanghai Institute of Applied Mathematics and Mechanics, Shanghai Key Laboratory of Mechanics in Energy Engineering, Shanghai University, Shanghai 200072, People's Republic of China*

We derive analytical expressions for the Young's modulus and the Poisson's ratio of the graphene nanoribbon, in which free edges are warped by the compressive edge stress. Our analytical formulas explicitly illustrate the reduction of the Young's modulus by the warped free edges, leading to the obvious width dependence for the Young's modulus of the graphene nanoribbon. The Poisson's ratio is also reduced by the warped free edges, and negative Poisson's ratio can be achieved in the graphene nanoribbon with an ultra-narrow width. These results are comparable with previous theoretical works. *Published by AIP Publishing.* https://doi.org/10.1063/1.5012562

## I. INTRODUCTION

Mechanical properties of graphene have been widely studied by previous experimental and theoretical works. Earlier experiments reported the Young's modulus of bulk graphite to be $1.06 \pm 0.02$ TPa.[1] In 2007, Frank *et al.*[2] measured the Young's modulus of the few-layer graphene, and found that the value is approximately at 0.5 TPa. In 2008, the Young's modulus of the monolayer graphene was measured to be $1 \pm 0.1$ TPa by Lee *et al.*[3] A huge number of theoretical works have also been devoted to the calculation of the Young's modulus for graphene. Liu *et al.*[4] performed density functional calculations to investigate the mechanical properties of graphene, and found that the Young's modulus of graphene is 1.05 TPa and the Poisson ratio is 0.186. Shokrieh and Rafiee[5] obtained the molecular potential energy and the Young's modulus of the graphene sheet is obtained by using the beam model. More discussions on the mechanical properties of graphene can be found in some recent review articles (see, for example, the study by Novoselov *et al.*[6] and Akinwande *et al.*[7])

As a result of its quasi-two-dimensional structure,[8,9] the free edges play an important role in various physical properties of the graphene nanoribbons (GNRs). In particular, the edge stress can cause strong effects on various mechanical properties. The origin of the edge stress is related to the under coordination of atoms at the free edge. The chemical bonds for the edge atoms will be different from bonds of the interior atoms, leading to compressive/tensile deformation of the free edge. In practice, the edge stress can be defined in analogy with the surface stress of a three-dimensional (3D) crystal.[10] At the free edges of the GNRs, the compressive edge stress can cause a transition from the planar configuration into the warped configuration,[10,11] which results from the buckling phenomenon[12] induced by the extremely high in-plane stiffness (35 GPa)[13] and ultrasmall bending modulus (1.44 eV).[14] It has been shown that the warping structure has strong effects on the elastic properties of GNRs. In 2016, Jiang and Park[15] found the negative Poisson's ratio in GNRs using an inclined plate model, where the value of the Poisson's ratio is governed by the interplay between the width and the warping amplitude of the edge. Furthermore, Jiang[16] also found that the three-dimensional warped structure with free edges will transform to the two-dimensional planner structure at the critical strain $\varepsilon = 0.7\%$. Recently, Zhang *et al.*[17] investigated the effect of the warped edge on the buckling process of GNRs.

In particular, the edge effects on the Young's modulus and the Poisson's ratio have also been investigated extensively. For instance, Reddy *et al.*[18] performed molecular dynamical (MD) simulations to demonstrate that the edges strongly affect the elastic properties of the graphene sheet when the width is less than 8 nm. Some studies have found that the Young's modulus of the GNR increases with increasing width, resulting from the free edges,[19,20] while others found an opposite width dependence for the Young's modulus.[21,22] The edge induced width dependence for the Poisson's ratio has also been investigated through some numerical methods.[20,22,23]

Overall, various numerical approaches have been adopted by most existing works to investigate the edge effects on the Young's modulus and the Poisson's ratio of the GNRs. An analytical study can explicitly disclose the relation between the free edge and the Young's modulus and the Poisson's ratio, which is still lacking. We thus provide an analytical derivation to reveal the direct relation between the warped free edge and the Young's modulus and the Poisson's ratio in the GNR.

In this paper, we derive the analytical expression for the Young's modulus and the Poisson's ratio of the GNRs, in which free edges are warped. The effect of the warped edge is considered analytically by considering the elastic energy of the warped configuration. The Young's modulus and the Poisson's ratio of the GNRs both increase monotonically with the increase of the width. Our analytical results are compared with the existing theoretical works.

## II. ELASTIC ENERGY DENSITY

Free edges in the graphene nanoribbons (GNRs) are warped by the compressive edge stress. The shape of the warped edge can be described by[10]

---

[a)]Author to whom correspondence should be addressed: jwjiang5918@hotmail.com

$$z(x, y) = A e^{-y/l_c} \sin(\pi x / \lambda_c), \quad (1)$$

where $A$ is the amplitude of the ripple, and $l_c$ is the penetration length. $\lambda_c$ is the wave length of the warping ripple. The z-axis is perpendicular to the graphene plane, while the x-axis is along the edge.

We can compute the strain energy of the warped edge, i.e., the edge energy. To do so, we consider a semi-infinite sheet structure in the region $-\infty < x < +\infty, 0 \leq y < +\infty$. The stresses are $\sigma_{13} = \sigma_{23} = \sigma_{33} = 0$ on the surface of the sheet. According to the Hooke's law, we have

$$\sigma_{33} = \frac{E}{(1+\nu)(1-2\nu)}[(1-\nu)u_{33} + \nu(u_{11}+u_{22})], \quad (2)$$

where $\nu$ is the Poisson's ratio and $E$ is the Young's modulus. From the boundary condition, $\sigma_{33} = 0$, we obtain

$$u_{33} = \frac{\nu}{(1-\nu)}(u_{11}+u_{22}). \quad (3)$$

From the expression of the warped configuration in Eq. (1), the strain in this structure can be obtained by $u_{ij} = \frac{1}{2}\frac{\partial z}{\partial x_j}\frac{\partial z}{\partial x_i}$, i.e.,

$$u_{11} = \frac{1}{2}\left(\frac{\partial z}{\partial x}\right)^2 \quad u_{22} = \frac{1}{2}\left(\frac{\partial z}{\partial y}\right)^2$$

$$u_{33} = \frac{\nu}{2(1-\nu)}\left[\left(\frac{\partial z}{\partial x}\right)^2 + \left(\frac{\partial z}{\partial y}\right)^2\right]$$

$$u_{12} = u_{21} = 0. \quad (4)$$

As a result, the warping induced strain energy is

$$U_1 = \frac{E}{2(1+\nu)}\left(u_{ik}^2 + \frac{\nu}{1-2\nu}u_{ll}^2\right). \quad (5)$$

Using the exact expression for each strain in Eq. (4), we obtain

$$U_1 = \frac{M}{8}\left[\left(\frac{\partial z}{\partial x}\right)^2 + \left(\frac{\partial z}{\partial y}\right)^2\right]^2, \quad (6)$$

where $M = E/(1-\nu^2)$.

The energy associated with the free edge is

$$U_2 = \tau_e u_{11} + \frac{1}{2}E_e u_{11}^2, \quad (7)$$

where $\tau_e$ is edge stress, which is determined by the bonding configuration of edge atoms and is thus a width independent constant. $E_e$ is the edge Young's modulus at the edge of the GNR.

Using the explicit expressions for the strain in Eq. (4), we obtain

$$U_2 = \frac{1}{2}\tau_e\left(\frac{\partial z(x,0)}{\partial x}\right)^2 + \frac{E_e}{8}\left(\frac{\partial z(x,0)}{\partial x}\right)^4. \quad (8)$$

The total energy in one periodic length $\lambda_c = 2\pi/k$ for the warped edge is

$$U_e = 2\left(\int_{x=0}^{2\pi/k} U_2 dx + \int_{x=0}^{2\pi/k}\int_{y=0}^{\infty} U_1 dx dy\right)$$

$$= \frac{A^2 \pi^2 \tau_e}{\lambda_c} + \frac{3A^4 \pi^4 E_e}{16\lambda_c^3} + \frac{A^4 M\left(3 + \frac{3l_c^4 \pi^4}{\lambda_c^4} + \frac{2l_c^2 \pi^2}{\lambda_c^2}\right)}{64 l_c^3}, \quad (9)$$

where the prefactor of 2 is to consider a pair of opposite free edges in the nanoribbon system.

The amplitude of the warped edge will decrease gradually by applying the tensile strain. The dependence of the warping amplitude on the tensile strain can be described by $A = A_0 \cos\left(\frac{\pi u_{11}}{2\varepsilon_c}\right)$, with $A_0 = 0.26$ nm as the amplitude of the warped edge without strain.[10] The warped edge will be fattened by tensile strain above the critical strain $\varepsilon_c = 0.7\%$.[16] Substituting this function of $A$ into Eq. (9), we obtain the strain dependence of the linear density for the edge energy

$$U_e = \frac{A_0^2 \pi^2 \tau_e \cos^2\left(\frac{\pi u_{11}}{2\varepsilon_c}\right)}{\lambda_c} + \frac{3A_0^4 \pi^4 E_e \cos^4\left(\frac{\pi u_{11}}{2\varepsilon_c}\right)}{16\lambda_c^3}$$

$$+ \frac{A_0^4 M\left(3 + \frac{3l_c^4 \pi^4}{\lambda_c^4} + \frac{2l_c^2 \pi^2}{\lambda_c^2}\right)\cos^4\left(\frac{\pi u_{11}}{2\varepsilon_c}\right)}{64 l_c^3}. \quad (10)$$

As a result, the edge energy per volume is

$$U_e = \frac{A_0^2 \pi^2 \tau_e \cos^2\left(\frac{\pi u_{11}}{2\varepsilon_c}\right)}{\lambda_c^2 W h} + \frac{3A_0^4 \pi^4 E_e \cos^4\left(\frac{\pi u_{11}}{2\varepsilon_c}\right)}{16\lambda_c^4 W h}$$

$$+ \frac{A_0^4 M\left(\frac{3}{\lambda_c} + \frac{3l_c^4 \pi^4}{\lambda_c^5} + \frac{2l_c^2 \pi^2}{\lambda_c^3}\right)\cos^4\left(\frac{\pi u_{11}}{2\varepsilon_c}\right)}{64 l_c^3 W h}, \quad (11)$$

where $w$ and $h$ are the width and the thickness of the system, respectively. We have introduced the quantity $M$

$$M = \frac{E_0}{1-\nu_0^2}. \quad (12)$$

In addition to the edge energy, there is the usual strain energy in graphene[24]

$$U_\epsilon = \mu\left(u_{ik} - \frac{1}{2}\delta_{ik}u_{ll}\right)^2 + \frac{1}{2}K u_{ll}^2, \quad (13)$$

where $K = \lambda + \mu$ is the bulk modulus and $\lambda$ and $\mu$ are the Lamé coefficients.

The total energy of the GNRs is the summation of the edge and the strain energy

$$U = U_e + U_\epsilon = \mu\left(u_{ik} - \frac{1}{2}\delta_{ik}u_{ll}\right)^2 + \frac{1}{2}Ku_{ll}^2$$

$$+ \frac{A_0^2\pi^2\tau_\varepsilon\cos^2\left(\frac{\pi u_{11}}{2\varepsilon_c}\right)}{\lambda_c^2 Wh} + \frac{3A_0^4\pi^4 E_e\cos^4\left(\frac{\pi u_{11}}{2\varepsilon_c}\right)}{16\lambda_c^4 Wh}$$

$$+ \frac{A_0^4 M\left(\frac{3}{\lambda_c} + \frac{3l_c^4\pi^4}{\lambda_c^5} + \frac{2l_c^2\pi^2}{\lambda_c^3}\right)\cos^4\left(\frac{\pi u_{11}}{2\varepsilon_c}\right)}{64l_c^3 Wh}. \quad (14)$$

## III. EDGE EFFECTS ON THE YOUNG'S MODULUS AND POISSON'S RATIO

The stress tensor can be derived from its definition, $\sigma_{ik} = \frac{\partial U}{\partial u_{ik}}$, which gives

$$\sigma_{ik} = Ku_{ll}\delta_{ik} + 2\mu\left(u_{ik} - \frac{1}{2}u_{ll}\delta_{ik}\right)$$

$$- \frac{1}{64\lambda_c^4 Wh\varepsilon_c^2}\left[32A_0^2\pi^4\lambda_c^2\tau_\varepsilon + 12A_0^4\pi^6 E_e + A_0^4 M\right.$$

$$\left.\times\left(\frac{3\pi^2\lambda_c^3}{l_c^3} + \frac{3l_c\pi^6}{\lambda_c} + \frac{2\lambda_c\pi^4}{l_c}\right)\right]\delta_{ik}u_{11}\delta_{ik}$$

$$= Ku_{ll}\delta_{ik} + 2\mu\left(u_{ik} - \frac{1}{2}u_{ll}\delta_{ik}\right) + Bu_{11}\delta_{ik}, \quad (15)$$

where we have introduced the parameter $B$

$$B = -\frac{1}{64\lambda_c^4 Wh\varepsilon_c^2}\left[32A_0^2\pi^4\lambda_c^2\tau_\varepsilon + 12A_0^4\pi^6 E_e + A_0^4 M\right.$$

$$\left.\times\left(\frac{3\pi^2\lambda_c^3}{l_c^3} + \frac{3l_c\pi^6}{\lambda_c} + \frac{2\lambda_c\pi^4}{l_c}\right)\right]. \quad (16)$$

We thus can obtain the strain tensor $u_{ik}$

$$u_{ik} = \frac{\delta_{ik}\sigma_{ll}}{2(2K+B)} + \frac{\sigma_{ik} - \frac{K\sigma_{ll} + B\sigma_{11}}{2K+B}\delta_{ik}}{2\mu}. \quad (17)$$

To calculate the Young's modulus and the Poisson's ratio, we stretch the graphene along the x-direction with a tension per area as $p$. The only nonzero component in the stress tension is $\sigma_{xx} = p$. According to Eq. (17), the nonzero components of the strain tensor are

$$u_{xx} = \left[\frac{1}{2(2K+B)} + \frac{K}{2\mu(2K+B)}\right]p$$

$$u_{yy} = \left[\frac{1}{2(2K+B)} - \frac{K+B}{2\mu(2K+B)}\right]p. \quad (18)$$

The Young's modulus and Poisson's ratio are thus obtained from their definitions

$$E = \frac{2(B+2K)\mu}{K+\mu}$$

$$\nu = \frac{B+K-\mu}{K+\mu}, \quad (19)$$

where the parameter $B$ describes the edge effect as defined in Eq. (16).

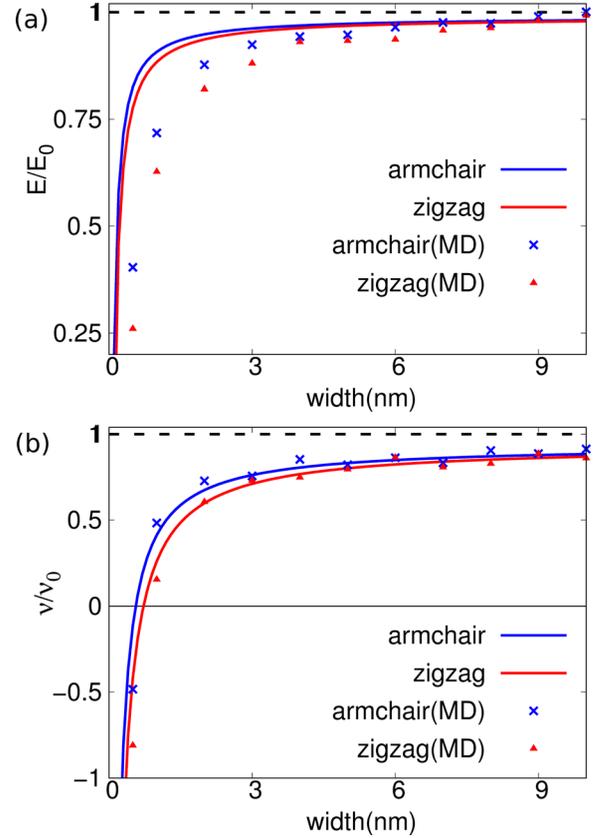

FIG. 1. The width dependence for the Young's modulus $E/E_0$ in (a) and Poisson's ratio $\nu/\nu_0$ in (b) for GNRs. Values are with respect to the value of pure graphene without free edges, i.e., $E_0 = 1.05$ TPa and $\nu_0 = 0.186$.[4] Lines are analytical predictions, while points are MD results.

Figure 1 shows the width dependence for the Young's modulus and the Poisson's ratio of the GNRs. We have used the following value for the parameters. The Young's modulus of the pure graphene without edge is $E_0 = 1.05$ TPa and the Poisson's ratio $\nu_0 = 0.186$.[4] The critical strain $\varepsilon_c = 0.7\%$ is from Jiang.[16] The other parameters for the edges of the graphene are from the original paper by Shenoy et al.[10] The initial amplitude $A_0 = 0.26$ nm, the penetration length $l_c = 2.3$ nm, and the wave length of the warping ripple $\lambda_c = 10$ nm are from the study by Shenoy et al.[10] The edge stresses are $\tau_e = 10.5$ eV and 20.5 eV for the armchair and zigzag edges, respectively. The edge Young's modulus values are $E_e = 112.6$ eV/nm and 147.2 eV/nm for the armchair and zigzag edges, respectively.

We also perform molecular dynamical (MD) simulations to compute numerically the width dependence for the Young's modulus and the Poisson's ratio. The MD simulations are performed using the publicly available simulation code LAMMPS.[25,26] The standard Newton equations of motion are integrated by using the velocity Verlet algorithm with a time step of 1 fs. The numerical results are in reasonable agreement with our analytical predictions, especially for GNRs of a large width. However, it should be noted that there are some deviations between the numerical results and the analytical predictions for ultra-narrow GNRs, which shall be attributed to the strong interplay between the two free edges that was not considered in the analytical derivation.

Figure 1(a) shows the width dependence for the Young's modulus. The Young's modulus for the zigzag GNRs is smaller than that of the armchair GNRs, but the difference is small. For both armchair and zigzag GNRs, the Young's modulus increases with increasing width, and will approach the value of the Young's modulus for the pure graphene without free edges. It has also been found by Zhao et al.[20] that the Young's modulus for both armchair and zigzag GNRs increase with the increase of width. Bu et al.[19] also found that the Young's modulus increases along the width for narrower GNRs. The work of Lu et al.[21] obtained an opposite result that the Young's modulus decreases with the increasing of GNR width.

Figure 1(b) shows the width dependence for the Poisson's ratio of the armchair and zigzag GNRs. The Poisson's ratio for both armchair and zigzag GNRs increase with the increase of the width, and will saturate at the value of the Poisson's ratio in pure graphene without free edges. There is a small difference in the Poisson's ratio between the armchair and zigzag GNRs, and the Poisson's ratio in the armchair GNR is slightly larger than that of the zigzag GNR. Our analytical results agree with the predictions by Georgantzinos et al.,[23] where the Poisson's ratio also increases with increasing width. Our results are different from the work by Wang et al.,[22] in which the Poisson's ratio of armchair (zigzag) GNRs decreases (increases) with the increasing width.

Figure 1(b) shows that, due to the edge effect, the Poisson's ratio can be negative for ultra-narrow GNRs, where the warped edge takes dominant effects. The warped edges will be effectively expanded during the flattening process under external stretching. This edge induced negative Poisson's ratio phenomenon was also found by Jiang and Park.[15]

We note that there are similar size effects of 3D materials.[27–30] In particular, for the 2D ribbon of width $w$, we can divide the system into three regions, including one interior region of size $w - 2l_c$ and two edge regions of size $l_c$. The penetration length $l_c$ can be regarded as the size of the edge region. Simple algebra gives the effective properties for the whole ribbon

$$D = D_0 - \frac{2l_c}{W}(D_0 - D_e), \quad (20)$$

where $D$ is the elastic properties like the Young's modulus and Poisson's ratio in this work, and $D_0$ and $D_e$ are the corresponding mechanical quantities for the interior and edge regions. We thus obtain a general formula for the width dependence of the effective Young's modulus and the Poisson's ratio. The width dependence is reflected by the parameter $B$ in Eq. (16), which is indeed inverse to the width of the ribbon and thus agrees with the general argument here. For 3D materials, there are similar general arguments to Eq. (20). As a result, there are similar size effects in the 3D structures. For example, Miller and Shenoy have shown that the elastic properties of nanosized 3D structural elements have a similar size dependence.[27]

## IV. CONCLUSION

To summarize, we have performed an analytical study for the effect of the warped edge on the mechanical properties for the graphene nanoribbon, including the Young's modulus and the Poisson's ratio. The warped edge effect is considered analytically by using the expression of the elastic energy of the warped configuration. We obtain the analytical expression disclosing the relation between the edge properties and the Young's modulus and Poisson's ratio. More specifically, the Young's modulus increases with the increase of width, and the Poisson's ratio also increases with increasing width. These results are comparable with previous works.

## ACKNOWLEDGMENTS

This work was supported by the Recruitment Program of Global Youth Experts of China, the National Natural Science Foundation of China (NSFC) under Grant No. 11504225, and the Innovation Program of Shanghai Municipal Education Commission under Grant No. 2017-01-07-00-09-E00019.